\documentclass[11pt]{article}
\DeclareMathAlphabet{\scr}{U}{rsfs}{m}{n}
\usepackage{latexsym}
\usepackage{epsfig}
\usepackage[mathscr]{eucal}
\usepackage{amsfonts}
\usepackage{amscd}
\usepackage{amsmath}
\usepackage{array}	
\usepackage{amssymb}
\usepackage{colordvi}
\usepackage{enumerate}
\usepackage{graphicx}
\usepackage{booktabs}
\usepackage[footnotesize]{caption}
\usepackage{fancyhdr} 
\usepackage{pdfpages}
\usepackage{slashed}
\usepackage{tabularx}
\usepackage{longtable}
\usepackage{array}
\usepackage{hyperref}
\usepackage{relsize}
\usepackage{color}
\usepackage[merge,numbers,compress]{natbib}

\hypersetup{
	colorlinks=true,        
	linkcolor=blue,          
	citecolor=black,        
	filecolor=red,      
	urlcolor=cyan           
}

\definecolor{ourpurple}{RGB}{145,0,140}
\definecolor{darkorange}{RGB}{225,100,0}
\definecolor{darkgreen}{RGB}{0,170,0}
\definecolor{darkgray}{RGB}{80,80,80}

\newcommand{\newc}{\newcommand}
\newc{\EW}{electroweak\;}
\newc{\DM}{dark matter\;}
\newc{\SM}{standard model\;}
\newc{\GEV}{\text{GeV}}
\newc{\KK}{Kaluza-Klein\;}
\newc{\ff}{fragmentation function\;}
\newc{\be}{\begin{equation}}
\newc{\ee}{\end{equation}}
\newc{\bi}{\begin{itemize}}
\newc{\ei}{\end{itemize}}
\newc{\benu}{\begin{enumerate}}
\newc{\eenu}{\end{enumerate}}
\newc{\bc}{\begin{center}}
\newc{\ec}{\end{center}}
\newc{\bfig}{\begin{figure}}
\newc{\efig}{\end{figure}}
\newc{\neutone}{\tilde{\chi}^0_1}

\DeclareMathSymbol{\Gamma}{\mathalpha}{letters}{"00}
\DeclareMathSymbol{\Delta}{\mathalpha}{letters}{"01}
\DeclareMathSymbol{\Theta}{\mathalpha}{letters}{"02}
\DeclareMathSymbol{\Lambda}{\mathalpha}{letters}{"03}
\DeclareMathSymbol{\Xi}{\mathalpha}{letters}{"04}
\DeclareMathSymbol{\Pi}{\mathalpha}{letters}{"05}
\DeclareMathSymbol{\Sigma}{\mathalpha}{letters}{"06}
\DeclareMathSymbol{\Upsilon}{\mathalpha}{letters}{"07}
\DeclareMathSymbol{\Phi}{\mathalpha}{letters}{"08}
\DeclareMathSymbol{\Psi}{\mathalpha}{letters}{"09}
\DeclareMathSymbol{\Omega}{\mathalpha}{letters}{"0A}
\DeclareMathSymbol{\varGamma}{\mathalpha}{operators}{"00}
\DeclareMathSymbol{\varDelta}{\mathalpha}{operators}{"01}
\DeclareMathSymbol{\varTheta}{\mathalpha}{operators}{"02}
\DeclareMathSymbol{\varLambda}{\mathalpha}{operators}{"03}
\DeclareMathSymbol{\varXi}{\mathalpha}{operators}{"04}
\DeclareMathSymbol{\varPi}{\mathalpha}{operators}{"05}
\DeclareMathSymbol{\varSigma}{\mathalpha}{operators}{"06}
\DeclareMathSymbol{\varUpsilon}{\mathalpha}{operators}{"07}
\DeclareMathSymbol{\varPhi}{\mathalpha}{operators}{"08}
\DeclareMathSymbol{\varPsi}{\mathalpha}{operators}{"09}
\DeclareMathSymbol{\varOmega}{\mathalpha}{operators}{"0A}

\newcommand{\D}{\mathrm{d}}

\begin{document}

\title{\hfill ~\\[-30mm]
\phantom{h} \hfill\mbox{\small TTK-15-23} 
\\[1cm]
\vspace{13mm}   \textbf{
Constraints on Majorana Dark Matter from the LHC and IceCube 
}}
\date{September 2015}
\author{
Jan Heisig$^{1}$\footnote{E-mail: \texttt{heisig@physik.rwth-aachen.de}}\;,
Michael Kr\"amer$^{1}$\footnote{E-mail:
  \texttt{mkraemer@physik.rwth-aachen.de}}\;,
Mathieu Pellen$^{1}$\footnote{E-mail: \texttt{pellen@physik.rwth-aachen.de}}\;,
Christopher Wiebusch$^{2}$\footnote{E-mail: \texttt{wiebusch@physik.rwth-aachen.de}}\\[9mm]
{\small\it 
$^1$Institute for Theoretical Particle Physics and Cosmology,}\\ {\small \it RWTH Aachen University, 52056 Aachen, Germany}\\[1mm]
{\small\it
$^2$ III. Physikalisches Institut B, RWTH Aachen University, 52056 Aachen, Germany}
}

\maketitle

\begin{abstract}
We consider a simplified model for Majorana fermion dark matter and explore 
constraints from direct, indirect and LHC collider searches. 
The dark matter is assumed to couple to the Standard Model through a vector 
mediator with axial-vector interactions. We provide detailed analyses of LHC mono-jet 
searches and IceCube limits on dark matter annihilation in the Sun. 
In particular, we develop a method for calculating limits on simplified WIMP 
dark matter models from public IceCube data, which are only available for a 
limited number of Standard Model final states. We demonstrate 
that LHC and IceCube searches for Majorana dark matter are complementary and 
derive new limits on the dark matter and mediator masses, including in addition 
constraints from LHC di-jet searches, direct detection and the dark matter relic density.  
\end{abstract}
\thispagestyle{empty}
\vfill
\newpage
\setcounter{page}{1}

\tableofcontents

\section{Introduction}

Weakly interacting massive particles (WIMPs) are attractive candidates for dark matter 
and are predicted by various extensions of the Standard Model (SM). Searches for WIMPs 
at the Large Hadron Collider (LHC) and through direct and indirect detection experiments 
are complementary, and probe different types of dark matter models and different regions 
of the model parameter space. 

To explore the nature of dark matter, and to be able to combine results from direct, indirect and 
collider searches, one may follow a more model-independent approach. So-called simplified 
models (see \textit{e.g.}\ \cite{Abdallah:2014hon,Malik:2014ggr,Abercrombie:2015gea} and 
references therein) describe dark matter and its experimental signatures with a minimal amount 
of new particles, interactions and model parameters. They thus allow us to explore the landscape 
of dark matter theories, and serve as a mediator between the experimental searches and more 
complete theories of dark matter. 

Many simplified models for dark matter have been proposed in the literature. Minimal mo\-dels 
describe dark matter by a single particle which interacts with the SM through a single mediator. 
We focus on a model with Majorana fermion dark matter and a vector mediator with axial-vector 
couplings to quarks. Such models predict spin-dependent WIMP-nucleon scattering cross 
sections and can thus be probed by IceCube in the search for dark matter annihilation in the 
Sun~\cite{Blumenthal:2014cwa,Catena:2015iea}. Previous work on dark matter models with 
\mbox{(axial-)}vector mediators~\cite{Buchmueller:2013dya, Buchmueller:2014yoa,
Lebedev:2014bba,Alves:2015pea,Chala:2015ama} has mainly focused 
on Dirac fermion dark matter and, in particular, has not analyzed dark matter limits from IceCube. 

The complementarity of IceCube and LHC searches for dark matter has been explored in 
Ref.~\cite{Blumenthal:2014cwa}, albeit in an effective field theory (EFT) approach, where 
the masses of all particles mediating the interaction between the SM and dark matter are 
assumed to be large compared to the energy scale of the process. The EFT limit is reliable 
to interpret the low-energy WIMP-nucleon interactions which are probed by direct detection 
experiments, but it may break down when analyzing dark matter searches with IceCube 
and with the LHC. While the capture of WIMPs in the Sun is well described by an EFT, 
the annihilation process is in general not. Indeed, as we will show, the IceCube limits 
are sensitive to details of the simplified model which cannot be described within the EFT\@. 
Furthermore, the EFT  may break down when probing dark matter production at the LHC: 
when the energy scale of the interaction is near or larger than the mass of the mediator, 
resonance effects become important, and the mediator has to be included in the particle 
spectrum of the model~\cite{Busoni:2013lha,Busoni:2014sya,Bai:2010hh,Fox:2011pm,
Fox:2011fx,Fox:2012ru,Goodman:2011jq,Busoni:2014haa,Buckley:2014fba,Harris:2014hga}. 

We thus provide a comprehensive analysis of collider, direct and indirect detection 
constraints on Majorana dark matter with vector mediators in the simplified model framework. 
The paper is organized as follows. The dark matter model is introduced in 
section~\ref{sec:model}. We derive limits on the model parameters from LHC 
searches in section~\ref{sec:collider}. In particular, we use the most recent results for 
mono-jet searches from both the ATLAS~\cite{Aad:2015zva} and 
CMS~\cite{Khachatryan:2014rra} collaborations. Section~\ref{sec:astro} addresses 
constraints from the dark matter relic density, from direct dark matter searches, and, 
in particular, searches for dark matter annihilation in the Sun with IceCube.
For a mediator that is lighter than the WIMP, annihilation into a pair of mediators can 
be dominant. In order to determine the IceCube model rejection factor in this region 
of parameter space we develop a method to estimate the limits for annihilation into 
two mediators on the basis of the limits for annihilation into the SM particles the 
mediator decays into. We validate this method by applying it to annihilation into top 
quark pairs for which we find very good agreement with the most recent public IceCube
dark matter annihilation limits \cite{Aartsen:2016exj} for that channel.
We finally combine the various constraints and 
derive new limits on the dark matter and mediator masses in section~\ref{sec:results}. 
Our conclusions are presented in section~\ref{sec:conclude}.

\section{A simplified model for Majorana fermion dark matter} \label{sec:model}

We focus on a minimal model with Majorana fermion dark matter, $\chi$, and a vector 
mediator, $V_\mu$, with axial-vector couplings to quarks, 
\begin{equation}
{\cal L} \supset g_\chi\, \bar \chi \gamma^\mu \gamma^5 \chi V_\mu 
+ g_q\, \bar q \gamma^\mu \gamma^5 q V_\mu\,.
\end{equation}
We assume universal couplings of the mediator to the SM quarks and neglect couplings 
to leptons.\footnote{%
The case of different couplings to up- and down-type quarks has been considered
{\it e.g.} in Ref.~\cite{Chala:2015ama}. However, as discussed in Ref.~\cite{Bell:2015sza}, 
gauge-invariance sets very tight constraints on the difference of these couplings.} 
Thus the model has only four independent parameters: the couplings of the mediator 
to the dark matter and the SM quarks, $g_\chi,$ and  $g_q$, and the  dark matter and 
mediator masses, $m_\chi$ and $M_V$, respectively. We require both couplings to be
$g_\chi, g_q < 4\pi$.  

The axial-vector interaction of Majorana dark matter with nuclei leads to spin-dependent 
scattering cross sections and contributes significantly to the dark matter capture rate in 
the Sun (see section~\ref{sec:IceCube}). Searching for dark matter annihilation in the Sun 
with the neutrino telescope IceCube can thus place strong limits on such models, which 
are competitive with direct detection bounds and with dark matter searches at the LHC. 

The width of the mediator, $\Gamma_V$, is determined by the particle masses and the 
couplings:
\be
 \label{eq:Vwidth}
\Gamma_V = \frac{M_V}{\pi} \left( \frac{g_\chi^2}{6} \left(1-4 \frac{m_{\chi}^2}{M_V^2} \right)^{3/2} + 
\,\mathlarger\sum^6_{i = 1} \frac{g_q^2}{4} \left( 1-4 \frac{m_{q_i}^2}{M_V^2} \right)^{3/2} \right) \,,
\ee
where $m_{q_i}$ is the mass of the SM quarks.

The dark matter cross section at the LHC and the WIMP-nucleon scattering cross section 
depend on the product of the couplings and on the width of the mediator, 
$\sqrt{g_q g_\chi}$ and $\Gamma_V$, respectively. We will thus present our results in 
terms of $\Gamma_V$ and $\sqrt{g_q g_\chi}$, rather than in terms of the individual 
couplings $g_q$ and $g_\chi$.

\begin{figure}[h!]
\centering
\setlength{\unitlength}{1\textwidth}
\begin{picture}(0.5,0.415)
  \put(0.0,-0.007){\includegraphics[width=0.49\textwidth]{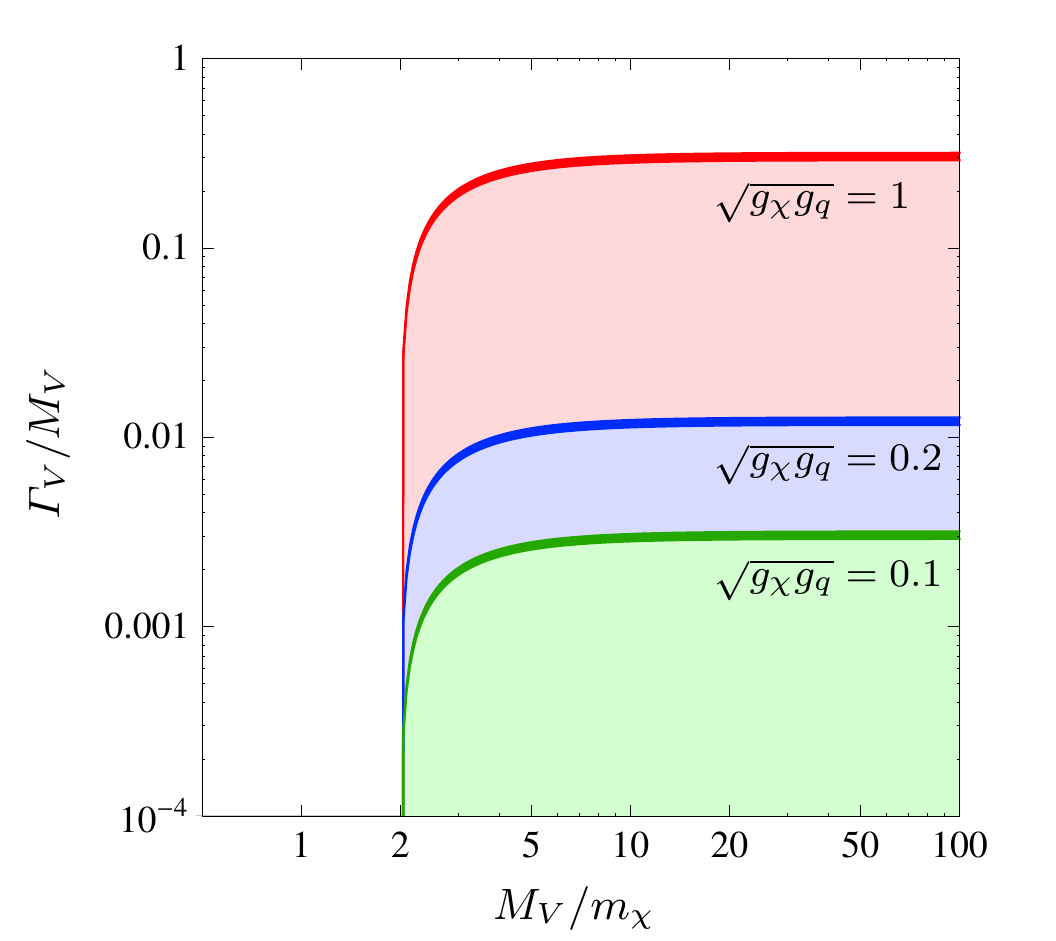}} 
\end{picture}
\caption{Non-accessible regions in the $M_V/m_\chi$-$\Gamma_V/ M_V$ plane for
$\sqrt{g_\chi g_q}=1$ (red shaded region), 0.2 (blue shaded region) and 0.1 (green
shaded region). The width of the bands marking the boundaries of the shaded regions 
indicates the explicit dependence on $m_\chi$: the upper and lower edge of the band 
correspond to the boundaries for $m_\chi=1\,\text{TeV}$ and $1\,\text{GeV}$, respectively.
}
\label{fig:allowedreg}
\end{figure}

The relation between $\Gamma_V$, $\sqrt{g_q g_\chi}$ and the individual couplings 
$g_q$ and $g_\chi$ provides some insight into the phenomenology of this model. 
In the region $M_V> 2m_\chi$, and for any given value of the product of the couplings, 
$\sqrt{g_q g_\chi}$, we encounter a minimal value for the mediator width, below which 
there is no solution for the individual couplings within the model. These regions are 
shown in Fig.~\ref{fig:allowedreg} in the $M_V/m_\chi$-$\Gamma_V/ M_V$ plane. 
The unaccessible regions only depend very mildly on $m_\chi$, as indicated by the 
width of the bands marking the boundaries of the shaded regions (here $m_\chi$ was 
varied between 1\,GeV and 1\,TeV). In the allowed part of the region where 
$M_V> 2m_\chi$ there exist two solutions for $g_q$ for any given $\sqrt{g_q g_\chi}$. 
To derive conservative limits on the model from di-jet production (see Sec.~\ref{sec:dijet}), 
we adopt the smaller value for $g_q$ in our analysis, unless this would cause $g_\chi>4\pi$ 
for a given $\sqrt{g_q g_\chi}$. 

In the EFT limit  $M_V\gg m_\chi$, where dark matter and SM quarks interact through 
a 4-fermion operator with coefficient $1/M_*^2=g_q g_\chi/M_V^2$, we find that 
$\Gamma_V/ M_V \gtrsim 0.3\, g_\chi g_q = 0.3\, (M_V/M_*)^2$. The LHC Run~I
data probe suppression scales $M_* \lesssim 1$\,TeV, see Sec.~\ref{sec:collider}. 
Thus, in the region $M_V \gg \sqrt{s}$ where the EFT is valid, $\Gamma_V/M_V$ 
is typically larger than one, inconsistent with a particle-like interpretation of the 
mediator (\textit{cf.} Ref.~\cite{Buchmueller:2013dya}).

\section{Collider limits} \label{sec:collider}

\subsection{Mono-jet limits}

Weakly interacting dark matter particles can be detected at the LHC through their 
associated production with jets, electroweak bosons or heavy quarks. The search for such 
signatures together with large missing transverse energy (MET) has been performed 
at the LHC Run~I and is 
one of the central goals of LHC Run~II~\cite{Abercrombie:2015gea}. In the following 
we will focus on signatures with mono-jets and MET as presented in \cite{Aad:2015zva,Khachatryan:2014rra}. Searches for electroweak gauge bosons with 
large MET are important in general, but provide weaker limits for the  dark matter 
model we consider, see e.g.\ \cite{Aad:2014tda,Khachatryan:2014rwa}.

To simulate the experimental signature for our model, we have generated events 
using \textsc{FeynRules}~2.1~\cite{Alloul:2013bka}, 
\textsc{Madgraph5\_aMC@NLO}~\cite{Alwall:2014hca} and 
\textsc{PYTHIA}~6~\cite{Sjostrand:2006za}, 
including QCD processes with one and two jets in the hard scattering. 
We have used  \textsc{DELPHES}~\cite{deFavereau:2013fsa} for the detector 
simulation and implemented the cuts employed in the ATLAS~\cite{Aad:2015zva} 
and CMS~\cite{Khachatryan:2014rra} mono-jet searches in an in-house program. 
With the observed and expected number of events provided by 
Refs.~\cite{Aad:2015zva, Khachatryan:2014rra} we are thus able to set exclusion 
limits at $95 \%$ confidence level (CL) on the different parameters of the model.

\begin{figure}[h!]
\centering
\setlength{\unitlength}{1\textwidth}
\begin{picture}(0.99,0.933)
  \put(0.0,-0.005){\includegraphics[width=0.98\textwidth]{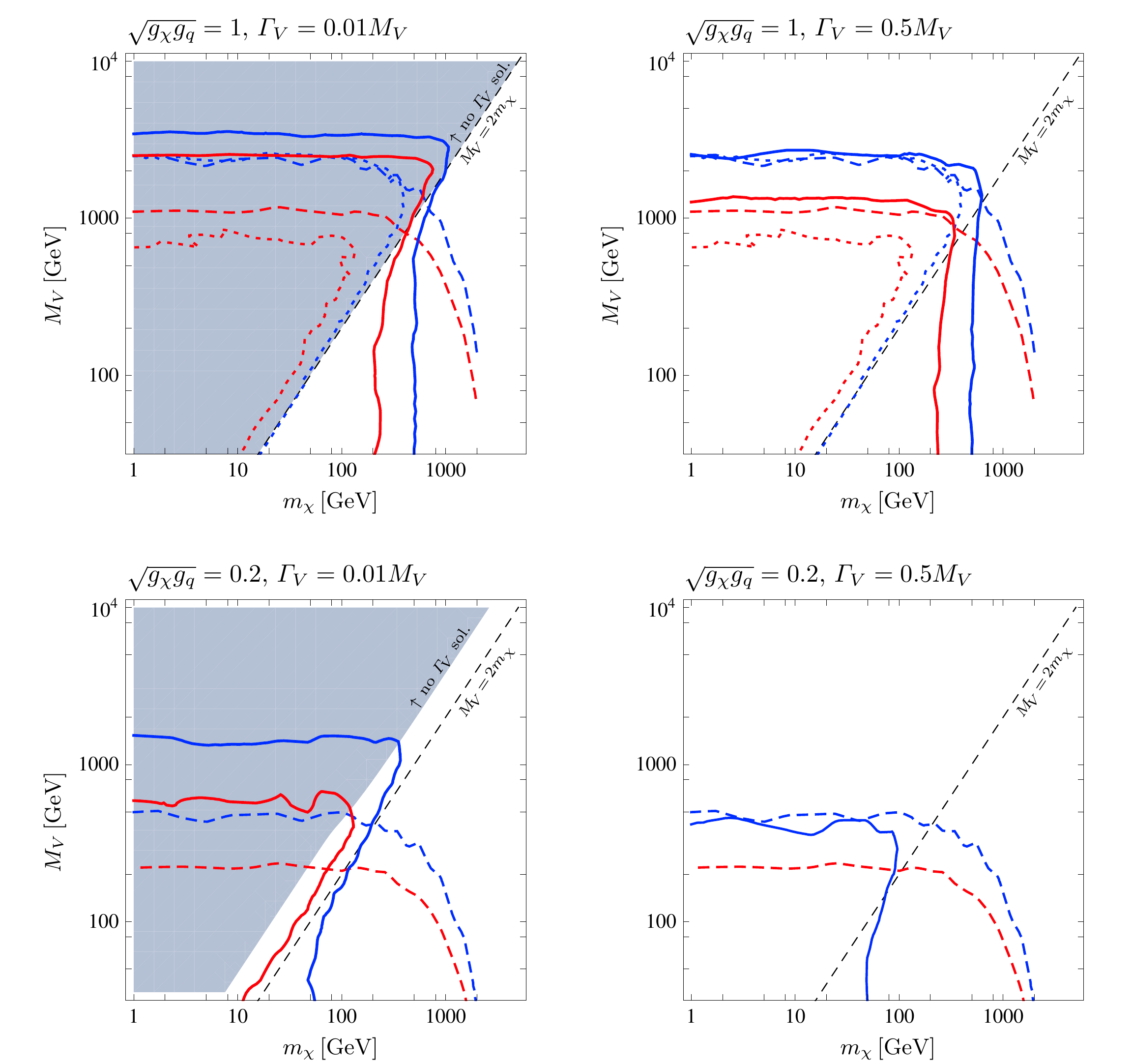}}
\end{picture}
\caption{
Lower exclusion limits in the $m_\chi$-$M_{V}$ plane at 95\% CL for the ATLAS (blue lines) 
and CMS (red lines) mono-jet searches. The limits for the simplified model (solid lines), 
for the EFT (dashed lines) and for the EFT applying the $Q$-truncation (dotted lines) 
are shown. Four slices of the parameter space: 
$\sqrt{g_\chi g_q}=1$ , $\Gamma_V=0.01 M_V$ (upper left panel),
$\sqrt{g_\chi g_q}=1$, $\Gamma_V=0.5 M_V$ (upper right panel),
$\sqrt{g_\chi g_q}=0.2$, $\Gamma_V=0.01 M_V$  (lower left panel) and
$\sqrt{g_\chi g_q}=0.2$, $\Gamma_V=0.5 M_V$  (lower right panel) are displayed.
The blue shaded region in the left panels represent the parameters space not 
allowing a consistent solution for the mediator width as a function of 
$ M_V,m_\chi,\sqrt{g_\chi g_q}$.
}
\label{fig:LHCres}
\end{figure}

We have also studied our model in the EFT limit, where the interaction is described 
by a higher-dimensional operator of the form 
$(g_\chi g_q/M_V^2) \, \bar{\chi} \gamma_\mu \gamma^5 \chi\,\bar{q}\gamma^\mu\gamma^5q$. 
As the EFT  is valid only for energy scales below the mediator mass, it has been 
proposed to restrict the momentum transfer in the $s$-channel, $Q < M_V$, when 
calculating cross sections at the LHC~\cite{Busoni:2014sya}. 

In Fig.~\ref{fig:LHCres} we show exclusion limits on the dark matter and mediator 
masses, for scenarios with a small or a large mediator width, $\Gamma_V/M_V=0.01$ 
and $0.5$, and small or large mediator couplings, $\sqrt{g_\chi g_q}=0.2$ and $1$, 
respectively. Note that we consider $\Gamma_V/M = 0.5$ a rather extreme benchmark case, which however is commonly adopted in the literature. For even larger $\Gamma_V/M$ the narrow-width approximation 
to the cross section calculation may not be reliable, and the interpretation of the mediator as a resonance becomes doubtful.
The results have been obtained for the ATLAS~\cite{Aad:2015zva} and 
CMS~\cite{Khachatryan:2014rra} mono-jet searches interpreted in terms of the simplified 
model, the EFT, and the EFT with a truncation $Q < M_V$. Note that not all combinations 
of the parameters $m_\chi, M_V, \Gamma_V$ and $\sqrt{g_\chi g_q}$ are viable, as 
discussed in section~\ref{sec:model}: for a small  width, $\Gamma_V/M_V = 0.01$, 
most of the region $M_V>2m_\chi$ is theoretically inconsistent. On the other hand, 
for a large width $\Gamma_V/M_V = 0.5$ the whole parameter region is allowed.

\begin{figure}[h!]
\centering
\setlength{\unitlength}{1\textwidth}
\begin{picture}(0.5,0.38)
  \put(0.0,-0.007){\includegraphics[width=0.5\textwidth]{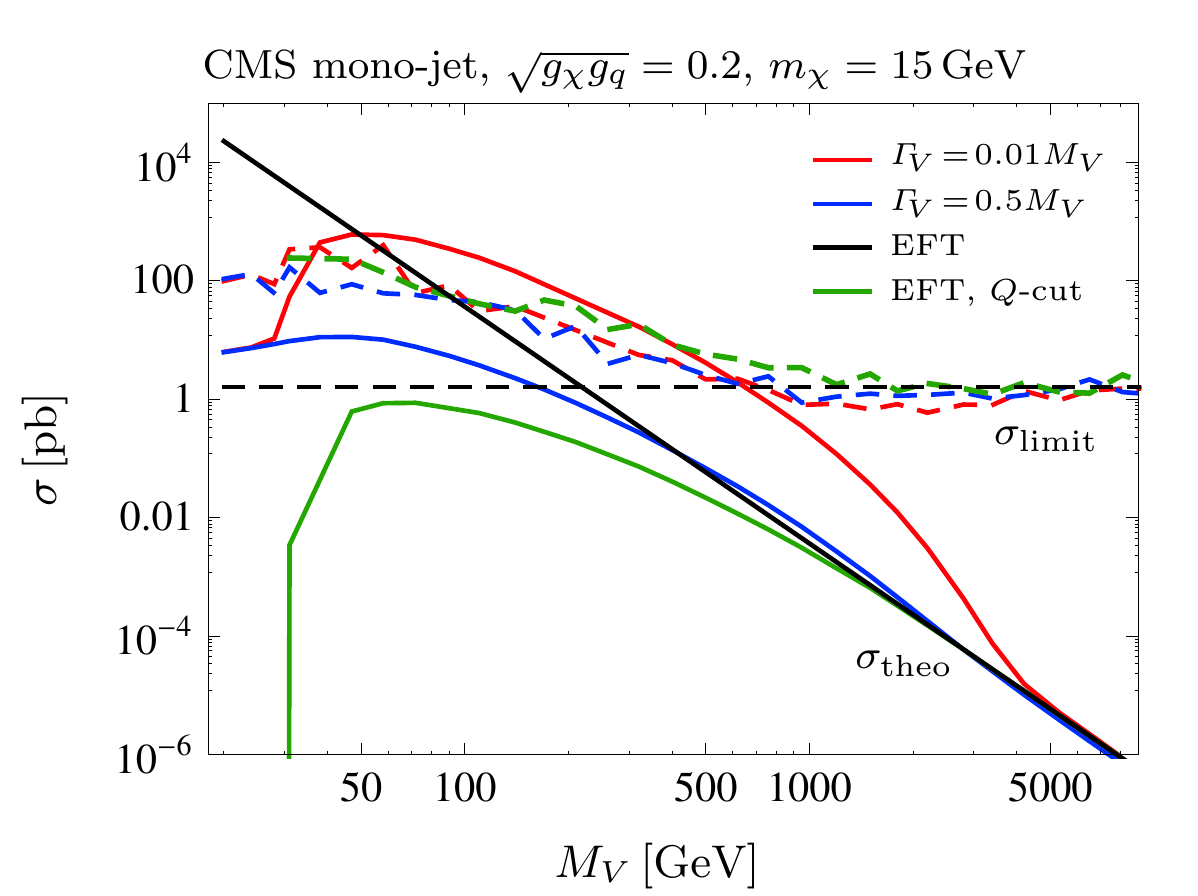}}
\end{picture}
\caption{
Theoretical prediction for the total production cross section (solid lines) and its observed 
upper limits (dashed lines) in the EFT limit with (green lines) and without (black lines) 
$Q$-truncation as well as in the simplified model with $\Gamma_V=0.01M_V$ (red lines) 
and $\Gamma_V=0.5M_V$ (blue lines). Example for $\sqrt{g_\chi g_q}=0.2$ and 
$m_\chi=15\,\GEV$ for the CMS analysis.}
\label{fig:xstheoLIM}
\end{figure}

We find large differences between the EFT and simplified model limits. For $M_V<2m_\chi$ 
the EFT limit (without $Q$-truncation) extents to much larger values of $m_\chi$ and 
drastically over-estimates the sensitivity in this region. By construction, this is not the 
case for the EFT limit employing the truncation on $Q$.
Furthermore, also for $M_V>2m_\chi$ we find significant differences. This is due
to the fact that $M_V$ lies in the range of accessible energies at the LHC.
The cross section can be greatly enhanced due to contributions from on-shell mediators. 
This leads to a larger sensitivity in the simplified model than in the EFT and is most 
pronounced for a small mediator width. However, for small $M_V$ -- smaller that the 
minimal MET of approximately $150$\,GeV required in the  search -- 
the cross section in the simplified model decreases again and the sensitivity becomes
weaker. This effect can be seen in the slice of parameter space with 
$\sqrt{g_\chi g_q}=0.2$ and $\Gamma_V=0.5M_V$ (lower right panel), 
where we cannot constrain the dark matter and mediator masses from the CMS analyses 
interpreted within the simplified model while within the EFT (without $Q$-truncation) a 
limit on $M_V$ around $200\,$GeV is obtained.

To illustrate the origin of the differences we present the cross-section predictions 
together with the observed upper CMS limits for the simplified model, the EFT 
and the EFT with truncation in Fig.~\ref{fig:xstheoLIM}. The results are shown 
exemplarily for $\sqrt{g_\chi g_q}=0.2$ and $m_\chi = 15$\,GeV, and for small 
and large widths, $\Gamma_V/M_V =0.01$ and $0.5$. The observed upper limits, 
which involve the observed number of events, the background expectation and the 
signal efficiencies, do not depend significantly on the mediator width, and are similar 
for the simplified model and the EFT with truncation. However, the cross section 
predictions from the simplified model and the EFT differ by orders of magnitude, 
in particular for small mediator widths. Furthermore, the simplified model cross section 
for $\sqrt{g_\chi g_q}=0.2$ and $\Gamma_V/M_V =0.5$ never exceeds the cross section 
limit, so that no constraint on the dark matter and mediator masses can be derived. Note 
that the slopes of the cross section limits and the cross section predictions are quite 
similar, so that small changes in the observed number of events, or small statistical 
fluctuations in the estimate of the efficiencies and cross sections, may have a visible 
impact on the limits.

We conclude that for our dark matter model, the EFT interpretation of the experimental 
LHC searches is only reliable if the mediator mass is larger than the accessible LHC 
partonic energies. However, as the current LHC limit on the suppression scale 
$M_*=M_V/\sqrt{g_\chi g_q}$ is of ${\cal O}(1\,{\rm TeV})$, an EFT interpretation 
would require extremely large couplings. On the other hand, 
the EFT limit with a truncation on $Q$ results in overly conservative limits 
which do not fully exploit the potential of the LHC searches.
We will thus use the limits obtained in the simplified model in the compilation 
of bounds presented in section~\ref{sec:results}.

\subsection{Di-jet limits} \label{sec:dijet}

Limits on our simplified model can also be derived from searches for the mediator 
particle in di-jet signatures. A detailed and comprehensive study of di-jet constraints 
has been presented  recently in Ref.~\cite{Chala:2015ama}, including results from 
UA2~\cite{Alitti:1993pn}, the Tevatron~\cite{Aaltonen:2008dn}, 
ATLAS~\cite{TheATLAScollaboration:2013gia,Aad:2014aqa} and 
CMS~\cite{Khachatryan:2015sja}. These searches can set bounds on a broad 
range of masses and are complementary to the signatures we have studied in 
this paper. We will  include the di-jet limits in the final discussion of the bounds 
on the model parameter space presented in section~\ref{sec:results}.

\section{Astrophysical and cosmological constraints} \label{sec:astro}

Models for dark matter are constrained by the relic density, indirect and 
direct searches. We have evaluated the relic density and the most recent 
IceCube limits on dark matter annihilation in the Sun \cite{Aartsen:2016exj}  
within our simplified model for Majorana dark matter, as discussed in detail 
below. We also briefly comment on direct dark matter searches, which provide 
further complementary constraints on our model. 

\subsection{Relic density} \label{sec:relic}

We have calculated the dark matter relic density for each point of the model 
parameter space using \textsc{FeynRules}~2.1~\cite{Alloul:2013bka}, 
\textsc{CalcHEP}~\cite{Belyaev:2012qa} and 
\textsc{micrOMEGAs}~4~\cite{Belanger:2014vza}. The calculation includes 
possible resonant effects, which are not taken into account when expanding in 
the dark matter velocity~\cite{Gondolo:1990dk}. We have however checked 
that the results obtained from the exact calculation within the simplified model 
match those obtained in the EFT limit for $M_V\gg m_\chi$.
To that end, we have calculated the annihilation cross section into quarks 
\cite{Zheng:2010js,Dreiner:2012xm} for the case of Majorana dark matter, 
\be
\begin{split}
\sigma_{qq} = \frac{g_q^2 g^2_{\chi} }{4 \pi} \frac{1}{ s M_V^4 \left( M_V^2 -s \right)^2 } \, \frac{\beta_q}{\beta_\chi}\,
\Big(\, &4 M_V^4 \left[ 28 m_q^2 m_{\chi}^2 - 4 s \left( m_{\chi}^2 + m_q^2 \right) + s^2\right] \\
&- \,96 M_V^2 m_{\chi}^2 m_q^2 s + 48 m_{\chi}^2 m_q^2 s^2 \Big)  \,,
\end{split}
\ee
where $\beta_{q,\chi} = \sqrt{1-4m^2_{q,\chi} / s}$ and $s$ is the center-of-mass energy.
Note that this annihilation cross section is helicity suppressed~\cite{Goldberg:1983nd,Go832}, 
and thus the main annihilation channels will be into  top and bottom quarks. 

In Fig.~6 (grey band) we show the parameter values of our model for which the relic 
density from thermal freeze-out agrees with the one measured by the Planck 
Collaboration~\cite{Ade:2013zuv}, $\Omega_\text{DM} h^2 = 0.1199$, within $\pm 10\%$. 
Assuming a standard cosmological history parameter points above this line can be 
considered excluded in the framework of our simplified model. Points below this line could 
be allowed by either requiring an additional component of dark matter or an additional 
(non-thermal) production mechanism. In this paper we assume 100\% of the (local) 
dark matter to be the considered WIMP candidate and hence do not rescale the limits from 
IceCube and LUX for points below the grey band. Note that in an extension of our simplified 
model co-annihilations or additional resonances may exist that could further weaken the relic 
density constraint while providing a similar phenomenology at the LHC and IceCube.

Finally, we mention that the unitarity of the $S$-matrix imposes constraints on the masses and couplings
in the dark matter model~\cite{Chala:2015ama,Griest:1989wd}. 
However, performing a definite analysis of unitarity constraints would require us to extend the 
simplified model framework, which is beyond the scope of this work.

\subsection{Limits from IceCube} \label{sec:IceCube}

If dark matter particles scatter in heavy astrophysical objects such as the Sun, they can lose enough 
energy to become gravitationally trapped inside the object. With the accumulation of 
dark matter, the annihilation rate can become large enough to lead to an equilibrium 
between dark matter capture and annihilation.

The evolution of the number of dark matter particles in the Sun, $N$, can be described 
by the Riccati differential equation \cite{Jungman:1995df}
\be
\label{eq:diffeqcap1}
\dot{N} = C_{\odot} - C_\text{A} N^2 - C_\text{E} N,
\ee
where $\dot{N}$ denotes the time derivative of $N$, 
$C_\odot$ is the capture rate of dark matter particles in the Sun, $C_\text{A} N^2 = 2 \Gamma_A$ 
is twice the dark matter annihilation rate and $C_\text{E} N$ is the evaporation rate, {\it i.e.}
the rate at which particles escape the Sun due to hard elastic scattering.
For dark matter particles with masses $m_\chi\gtrsim10\,\GEV$, the evaporation term can be 
neglected~\cite{Griest:1986yu}, allowing for a simple solution of Eq.~(\ref{eq:diffeqcap1}): 
\be
\label{eq:tanhsq}
C_\text{A}N^2  = C_\odot \tanh^2 \left(\sqrt{C_{\odot} C_A} \, t\right) \,.
\ee
For large times,  $\sqrt{C_\odot C_A} \, t \gg 1$, the tanh-term in Eq.~\eqref{eq:tanhsq} approaches 
one, and WIMP annihilation and capture are in equilibrium, \emph{i.e.}~$C_\odot = 2 \Gamma_A = C_A N^2$. 
This implies $\dot{N} = 0$. Hence in equilibrium the annihilation rate does not depend on the 
annihilation cross section, but only on the capture rate, which in turn is determined by the 
elastic WIMP scattering cross section. We shall analyze the equilibrium condition within our 
model in more detail in Sec.~\ref{sec:EQcond}.

\medskip 

Through a measurement of the neutrino flux, neutrino telescopes are sensitive to the WIMP 
annihilation rate. We consider data from the IceCube Neutrino Observatory~\cite{Aartsen:2012kia,Aartsen:2016exj} 
taken during 317 days in the years 2011 and 2012. No significant excess over background has 
been observed and these measurements can thus be used to set limits on possible dark matter signals.

The search has been interpreted in terms of limits on the spin-dependent WIMP-proton scattering 
cross section. The WIMP-nucleon scattering cross section in our model is 
\be
\begin{split}
\sigma_\text{SD}^{(N)}  &= \frac{12\mu_{N\chi}^2\,g_\chi^2}{\pi M_V^4} \left(
\sum_{q=u,d,s} g_q\, \Delta_q^{(N)}\right)^2\\
&\simeq1.8\times10^{-40}\,\text{cm}^2 
\left(\frac{\mu_{N\chi}}{1\,\GEV}\right)^2
\left(\frac{g_\chi g_q}{1}\right)^2
\left(\frac{1 \,\text{TeV}}{M_V}\right)^4 \,,
\label{eq:sigmaSDm}
\end{split}
\ee
where $\mu_{N\chi}=m_\chi m_N/(m_\chi + m_N)$ is the reduced WIMP-nucleon mass and 
$N=p$ for WIMP-proton scattering.\footnote{%
The scattering cross section for Majorana fermion dark matter given
in Eq.~\eqref{eq:sigmaSDm} is larger than the one for Dirac fermion 
dark matter by a factor of four (see \emph{e.g.}~Ref.~\cite{Agrawal:2010fh}).
This is in contradiction with the result quoted in Ref.~\cite{Savage:2015xta}, 
which is a factor of eight smaller than our result.}
In the second line of Eq.~\eqref{eq:sigmaSDm} we have used the fact that $g_q$ is universal 
and included the numerical values for the nucleon form factors
$\Delta_u^{(p)}=\Delta_d^{(n)}=0.85$, $\Delta_u^{(n)}=\Delta_d^{(p)}=-0.42$
and $\Delta_s^{(p)}=\Delta_s^{(n)}=-0.08$~\cite{Cheng:2012qr}. 
Considering universal couplings and neglecting the small mass difference 
between the proton and the neutron, the WIMP-neutron scattering cross section probed 
in direct detection experiments (see section~\ref{sec:LUX}) is equal to the WIMP-proton 
cross section.

The dark matter interpretation of the IceCube searches initially relied on two scenarios: 
annihilation into $b\bar b$ and $W^+W^-$ \cite{Aartsen:2012kia}.
Recently, an improved interpretation \cite{Aartsen:2016exj} has been performed which 
includes more scenarios, in particular annihilation into $t\bar t$.
In the model considered here, annihilation into $t\bar t$, $b\bar b$ 
and $VV$ is dominant (\textit{cf.}~Fig.~\ref{fig:cosmores}).
Therefore, in Sec.~\ref{sec:ttlims} we will first estimate the limit 
on the spin-dependent WIMP-proton scattering cross section for annihilation 
into $t\bar t$ based on the limits for the annihilation into $W^+W^-$ and $b\bar b$. 
The excellent agreement between our estimate of the  $t\bar t$ limits and the results 
presented in Ref.~\cite{Aartsen:2016exj} is considered as evidence for the accuracy 
of the conversion method. We then apply the same method to estimate limits
on the annihilation into $VV$ in Sec.~\ref{sec:VVlims}. For the exclusion limits
on the parameter space of our model we conservatively take into account only the 
most constraining channel and compute the model rejection factor, $\mu$, via
\be
\mu(m_\chi,M_V)=\sigma_\text{SD}^{(p)}(m_\chi,M_V)\times\max\!\left(  
\frac{R_{t \bar t}(m_\chi,M_V)}{\sigma_{t \bar t}^\text{UL}(m_\chi)}  
,\,\frac{R_{b\bar b}(m_\chi,M_V)}{\sigma_{b\bar b}^\text{UL}(m_\chi)} 
,\,\frac{R_{VV}(m_\chi,M_V)}{\sigma_{VV}^\text{UL}(m_\chi,M_V)}  
\right)\,,
\label{eq:mucomb}
\ee
where $R_{i}$ is the contribution of the channel $i$ to the annihilation cross section and 
$\sigma_{i}^\text{UL}$ is the corresponding upper limit on the spin-dependent
WIMP-proton scattering cross section. Note that we display the dependence of the various 
factors on the dark matter and mediator masses, but  not the dependence on the couplings 
$g_q,g_\chi$. A point in parameter space is excluded if $\mu\ge1$.

\subsubsection{Limit on $\sigma_\text{SD}$ for annihilation into $t\bar t$} \label{sec:ttlims}

In this subsection we will derive a limit on the spin-dependent WIMP-proton 
scattering cross section for dark matter annihilating to 100\% into $t\bar t$.
Our procedure uses the limits of capturing rates $\Gamma_\text{A}$
for annihilation into $W^+W^-$ and $b\bar b$ as input.
As WIMPs are practically at rest inside the Sun, the $W$ bosons (or $b$ quarks) 
that are produced directly in WIMP annihilation have a well-defined energy given by 
the WIMP mass $m_\chi$. Therefore we can interpret the limits on $\Gamma_\text{A}$ 
as limits on the annihilation rate of the $W$-(or $b$-)pairs with the energy $E^{W/b}=m_\chi$, 
regardless of the actual annihilation process.\footnote{%
Note that this is only true for the limit on $\Gamma_\text{A}$, but not for the limit on 
$\sigma_\text{SD}$ because $\sigma_\text{SD}$ depends explicitly on $m_\chi$ through the 
capture efficiencies. Furthermore, the limit on $\Gamma_\text{A}$ is independent of the 
correlation between the two $W$ bosons (or $b$ quarks) of one annihilation. 
The fraction of events where multiple neutrinos arising from $W$ bosons (or $b$ quarks) 
of the same annihilation simultaneously interact in the detector can safely be neglected.}
We exploit this fact and consider the energy spectrum of the $W$ bosons and $b$ quarks 
arising from the decay of the top quarks from WIMP annihilation into $t\bar t$. 
Since the interaction time of the top quark in the Sun is much larger than its lifetime, 
no energy loss is expected before it decays into an on-shell $W$-boson and  $b$-quark.
As a next step we calculate the probability distributions $P(E_i|E_t) $ of the energy 
of a final state particle $i$ for a given top energy -- and hence WIMP mass -- by simulating 
the annihilation process $\chi\chi\to t \bar t \to W^+ W^- b \bar b$ with 
\textsc{Madgraph5\_aMC@NLO}. The probability distributions are normalized to one.
Based on these distributions we can calculate the resulting limit on $\Gamma_\text{A}^{t \bar t}$ 
by a weighted average of the limits for each relevant final state $x=W^+,W^-,b,\bar b$:
\be
\frac{1}{\Gamma_\text{A}^{t \bar t}(E_t)} = 
\int_0^\infty \D E_W \,  
\frac{P(E_W|E_t) }{\Gamma_\text{A}^{W^+W^-}(E_W)}
+
\int_0^\infty  \D E_b \,  
\frac{P(E_b|E_t) }{\Gamma_\text{A}^{b\bar{b}}(E_b)}\,.
\label{eq:avGamma1}
\ee
In order to evaluate $\Gamma_\text{A}^{W^+W^-\!/b\bar b}$ for arbitrary values of the energy we 
interpolate the limits linearly in $E^{W/b}$ between the values given in  Ref.~\cite{Aartsen:2016exj} 
on a double logarithmic scale. Note that the contribution from the $b$ quarks in Eq.~\eqref{eq:avGamma1} 
is sub-leading. The correction is below $10\%$ (for large WIMP masses and decreases to below 
$1\%$ towards small masses).

\begin{figure}[h!]
\centering
\setlength{\unitlength}{1\textwidth}
\begin{picture}(0.46,0.35)
  \put(0.0,-0.007){\includegraphics[width=0.47\textwidth]{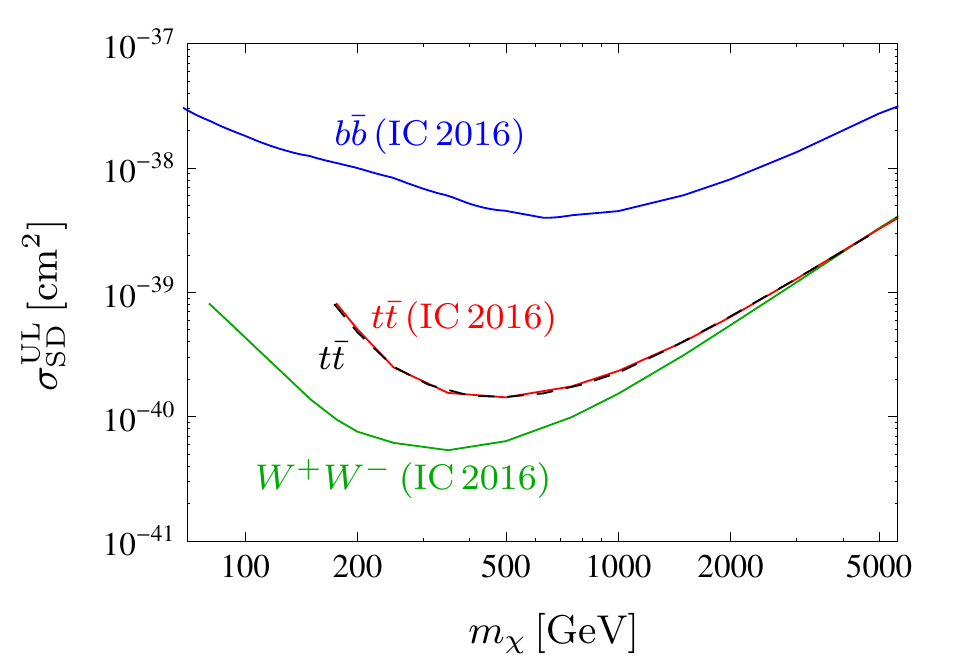}}
\end{picture}
\caption{
90\% CL upper limits on the spin-dependent WIMP-proton scattering cross section, 
$\sigma^\text{UL}_{\text{SD}}$, for 100\% annihilation into $t\bar t$ (red solid line), 
$b\bar b$ (blue dotted line) and $W^+W^-$ (green dashed line) from the IceCube 
Neutrino Observatory~\cite{Aartsen:2016exj}. The black dashed line shows the 
$t\bar t$ limit we derived from the $b\bar b$ and $W^+W^-$ data according to the 
method described in Sec.~\ref{sec:ttlims}.
}
\label{fig:xsSDlimICcomp}
\end{figure}

We converted the limits on the spin-dependent WIMP-proton scattering cross section
given in Ref.~\cite{Aartsen:2016exj} into limits on the annihilation rate (and vice versa) by
\be
\label{eq:conv}
\Gamma_\text{A}^\text{UL}(m_\chi)=\frac12 K_\text{SD}(m_\chi) \,\sigma_\text{SD}^\text{UL}(m_\chi)\,,
\ee
where $K_\text{SD}$ is the (channel independent) capture efficiency for the spin-dependent 
part of the scattering. We take $K_\text{SD}$ from Ref.~\cite{Aartsen:2012kia}
where it was derived following the method of Ref.~\cite{Wikstrom:2009kw} adopting the 
Standard Solar Model BS05(OP)~\cite{Bahcall:2004pz} as implemented in 
\textsc{DarkSUSY}~\cite{Gondolo:2004sc}. 

The result is presented in Fig.~\ref{fig:xsSDlimICcomp}, where we also show the limit on 
100\% annihilation into $W^+W^-$, $b\bar b$  and $ t\bar t$ from Ref.~\cite{Aartsen:2016exj}.
The good agreement of our cross section limits for ${t \bar t}$ with Ref.~\cite{Aartsen:2016exj} 
demonstrates the validity of our calculation. In addition, as a consistency 
check of our procedure, we have converted the limit on the annihilation rate into a limit on the resulting 
muon flux in the detector, $\Phi^\text{UL}_\mu$, using the conversion functions provided in 
Ref.~\cite{Wikstrom:2009kw}. As expected, we found that $\Phi^\text{UL}_\mu$ from the 
annihilation into $t\bar{t}$ is always in between the limits obtained for annihilation into 
$W^+W^-$ and $b\bar{b}$ as presented in Ref.~\cite{Aartsen:2016exj}.

\subsubsection{Limit on $\sigma_\text{SD}$ for annihilation into $VV$} \label{sec:VVlims}

In this subsection we will derive limits on the spin-dependent WIMP-proton scattering 
cross section for dark matter annihilating to 100\% into mediator pairs $VV$ 
with the method described in Sec.~\ref{sec:VVlims}. Annihilation into $VV$ takes 
place only for $M_V<m_\chi$, where the mediator decays solely into quarks. As we 
consider a universal coupling to all quarks, the corresponding branching
ratios are simply determined by the accessible phase space. Here we
only take into account neutrinos arising from the decay of the mediator 
into bottom and top quarks. In order to justify this choice we computed
the differential neutrino spectra $\D n_\nu/\D E_\nu$ for annihilation
into all quark flavors with \textsc{WimpSim}~3~\cite{Blennow:2007tw}.\footnote{%
The program package \textsc{WimpSim} is linked to 
\textsc{PYTHIA}~6~\cite{Sjostrand:2006za} for the simulation of dark matter
annihilation in the Sun, \textsc{Nusigma}~\cite{nusigma} for the simulation of 
neutrino-nucleon interactions and to \textsc{DarkSUSY}~\cite{Gondolo:2004sc} 
for the implementation of the Sun's density profile. We take the neutrino oscillation 
parameters from Ref.~\cite{Tortola:2012te}.} 
As expected, the neutrino fluxes spectra for light flavor quarks $d,u,s$ are much 
softer than for bottom and top quarks and can be safely neglected for the
derivation of the limits. The neutrino 
flux for annihilation into charm quarks is weaker than that for $b$-quarks by 
a factor of 3 to 10 (and much weaker than for top quarks)
in the relevant energy range and is hence subdominant.
We simulate the annihilation process $\chi\chi\to VV$ and the subsequent decays 
$V\to q\bar q$ with \textsc{Madgraph5\_aMC@NLO} and determine the  probability 
distributions $P(E_q|E_V) $ of the energy of a final state quark $q=b,t$, which are hence 
normalized to $2\times\text{BR}(V\to q\bar q)$. We calculate  the resulting limit on 
$\Gamma_\text{A}^{VV}$ analogous to Eq.~\eqref{eq:avGamma1} through:
\be
\frac{1}{\Gamma_\text{A}^{VV}(E_V)} = 
\int_0^\infty  \D E_t \,  
\frac{P(E_t|E_V) }{\Gamma_\text{A}^{t \bar t}(E_t)}
+
\int_0^\infty  \D E_b \,  
\frac{P(E_b|E_V) }{\Gamma_\text{A}^{b\bar{b}}(E_b)} \,.
\label{eq:avGamma}
\ee

\begin{figure}[h!]
\centering
\setlength{\unitlength}{1\textwidth}
\begin{picture}(0.98,0.35)
  \put(0.0,-0.007){\includegraphics[width=1\textwidth]{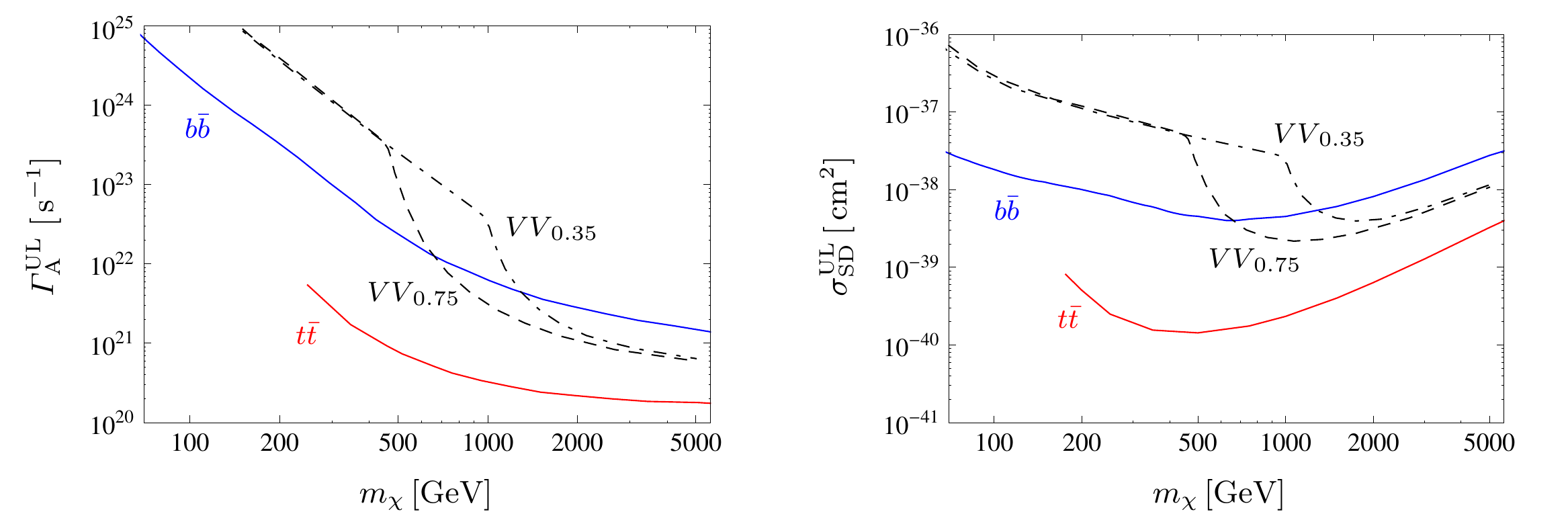}}
\end{picture}
\caption{
90\% CL upper limits on the annihilation rate $\Gamma_\text{A}$ and the spin-dependent WIMP-proton 
scattering cross section $\sigma_{\text{SD}}$ for 100\% annihilation into $VV$
(black lines) derived from a reinterpretation of the data from the IceCube Neutrino 
Observatory~\cite{Aartsen:2016exj}. We show two slices regarding the mediator mass, 
$M_V/m_\chi=0.75$ (dashed line) and $M_V/m_\chi=0.35$ (dot-dashed line).
For comparison we also show the limits for 100\% 
annihilation into $b\bar b$ (blue solid line) and $W^+W^-$ (red solid line) taken 
from Ref.~\cite{Aartsen:2016exj}.}
\label{fig:IClimitsVV}
\end{figure}

As the resulting limit depends on the WIMP mass and the mediator mass
we scan the corresponding two-dimensional grid. 
In Fig.~\ref{fig:IClimitsVV} we present the limits on $\Gamma_\text{A}$ (left panel) and 
$\sigma_\text{SD}^{VV}$ (right panel). The limits are considerably weaker than for 
annihilation into a pair of tops. As the limits depend both on $m_\chi$ and $M_V$ 
we show two slices in the parameter space $M_V/m_\chi=0.75$ and $M_V/m_\chi=0.35$, 
respectively. The main difference between these two slices is due to the opening of the 
mediator decay into top quarks for $M_V>2m_t$, which greatly enhances the sensitivity.
\medskip

\begin{figure}[h!]
\centering
\setlength{\unitlength}{1\textwidth}
\begin{picture}(0.99,0.933)
  \put(0.0,-0.007){\includegraphics[width=0.98\textwidth]{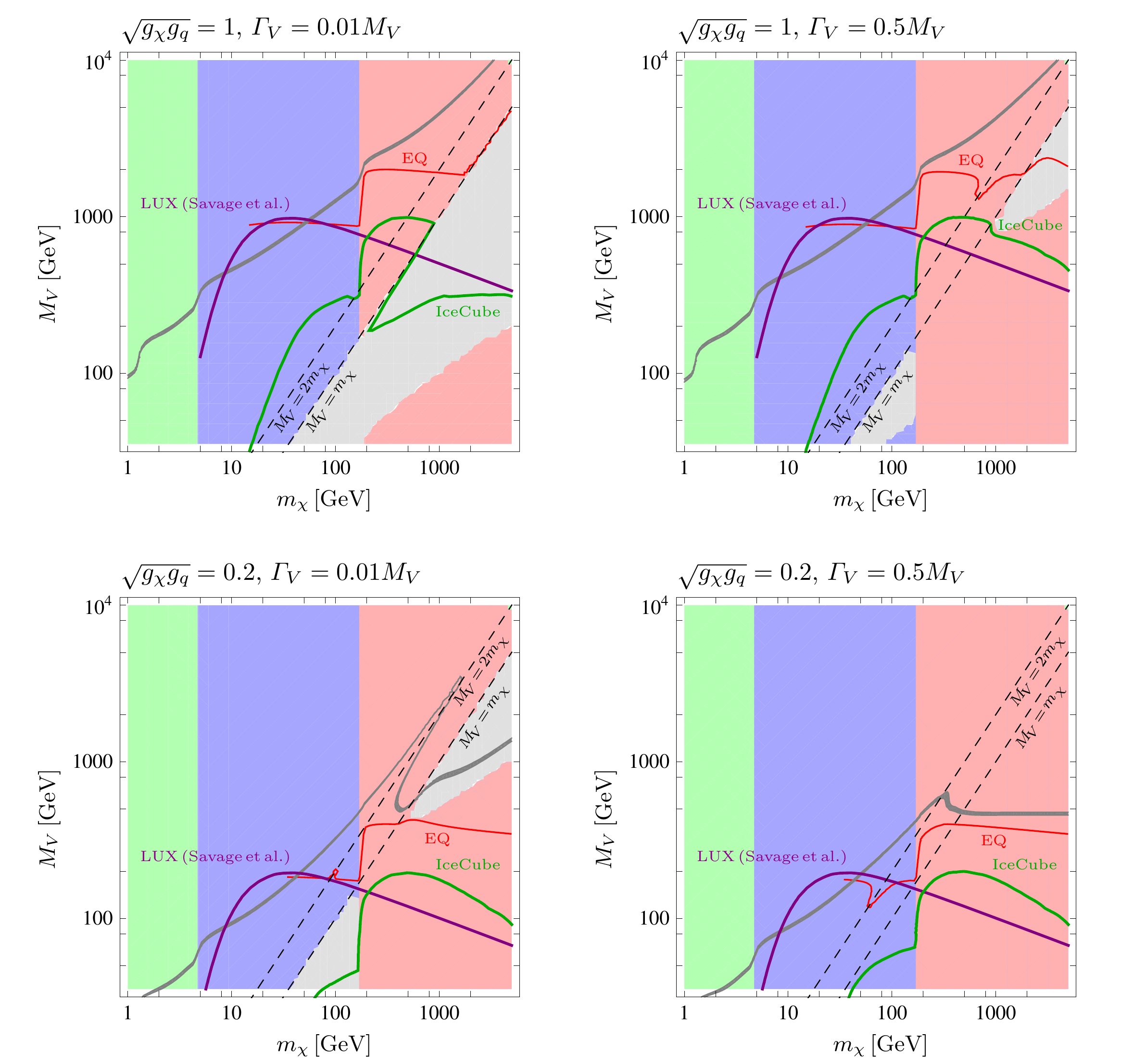}}
\end{picture}
\caption{
Astrophysical and cosmological quantities in the $m_\chi$-$M_{V}$ plane
in four slices of the considered parameter space: 
$\sqrt{g_\chi g_q}=1$, $\Gamma_V=0.01 M_V$ (upper left panel),
$\sqrt{g_\chi g_q}=1$, $\Gamma_V=0.5 M_V$ (upper right panel),
$\sqrt{g_\chi g_q}=0.2$, $\Gamma_V=0.01 M_V$  (lower left panel) and
$\sqrt{g_\chi g_q}=0.2$, $\Gamma_V=0.5 M_V$  (lower right panel).
The shaded regions denote the dominant annihilation channel, in red,
blue, green and grey we denote dominant annihilation into $t\bar t$,
$b\bar b$, light flavor quarks and two mediators, respectively.
The 90\% CL lower exclusion limits from the IceCube Neutrino Observatory 
(green line) and from LUX derived from the limits presented
in Ref.~\cite{Savage:2015xta} (purple line) are also displayed.
The dark grey shaded band denotes the region where the relic density 
matches the dark matter density within $\pm10\%$. In the region below 
the red thin line the equilibrium condition is fulfilled, \textit{i.e.}
$\sqrt{C_\odot C_A} \, t_\odot > 3$.
}
\label{fig:cosmores}
\end{figure}

The resulting IceCube limits on the model parameter space considering the annihilation channels $b\bar b$, 
$t\bar t$ and $VV$ are shown in Fig.~\ref{fig:cosmores}. The sensitivity to annihilation into $VV$ is significantly
weaker than for $t \bar t$ and -- below the $t \bar t$ threshold -- for $b \bar b$ final states. This causes a 
drop in the limit on $M_V$ for regions where annihilation into $VV$ is dominant (light grey areas).

\subsubsection{Equilibrium condition for capturing and annihilation in the Sun}
\label{sec:EQcond}

As discussed at the beginning of Sec.~\ref{sec:IceCube}, for large times, $\sqrt{C_\odot C_A}\, t \gg 1$,  
dark matter matter annihilation and capture in the Sun are in equilibrium, and the limits on the 
annihilation rate can directly be translated into limits on the elastic WIMP-nucleon scattering. 
Assuming that the Sun has been collecting dark matter during its entire lifetime, 
$t = t_\odot \simeq 1.5 \times 10^{17} \, \text{s}$, the equilibrium condition can 
approximately be expressed by~\cite{Jungman:1995df}
\be
\label{eq:equilcon}
\sqrt{C_\odot C_A} \, t_\odot \,\simeq \,330 \left( \frac{C_\odot}{\text{s}^{-1}} \right)^{1/2}
 \left( \frac{\langle \sigma_\text{A} v \rangle}{\text{cm}^3 \, \text{s}^{-1}} \right)^{1/2}
  \left( \frac{m_\chi}{10 \,\GEV} \right)^{3/4}\gg 1\,.
\ee
In practice $\sqrt{C_\odot C_A} \, t_\odot \gtrsim 3$ is already enough to obtain an error of 
less than a percent on $C_\odot$. In order to estimate $\sqrt{C_\odot C_A} \, t_\odot$ for our model, we compute 
$\langle \sigma_\text{A} v \rangle$ with \textsc{micrOMEGAs}. The 
capture rate, $C_\odot$, can be deduced from the WIMP-nucleon scattering cross section 
using the capture efficiency (see Sec.~\ref{sec:ttlims})
\begin{equation}
C_\odot =  K_\text{SD} (m_\chi) \sigma_\text{SD}\,.
\end{equation}
In Fig.~\ref{fig:cosmores} we show the contours $\sqrt{C_\odot C_A} \, t_\odot=3$ 
(labeled ``EQ") in the $m_\chi$-$M_V$ plane. Below this line the equilibrium condition
is fulfilled and the interpretation of the IceCube measurement in terms of the elastic 
scattering cross section is justified. A similar conclusion was put forward for the 
limits set in Ref.~\cite{Blumenthal:2014cwa}, where $\langle \sigma_\text{A} v \rangle$ 
was computed within the EFT\@. However, the fact that the equilibrium condition holds 
in the EFT limit does not imply that it should hold in the simplified model description, 
as the thermally averaged cross section, $\langle \sigma_\text{A} v \rangle$, can be 
smaller in the latter case.

\subsection{Limits from direct detection} \label{sec:LUX}

Although providing much stronger limits on the spin-independent WIMP-nucleon 
scattering cross section, direct detection experiments are also capable of imposing 
limits on spin-dependent scattering.
In the relevant part of the parameter space the strongest constraints are set by the 
LUX experiment~\cite{Akerib:2013tjd}.
As an up-to-date dedicated analysis for spin-dependent scattering has not been 
provided by the LUX Collaboration, several authors have reinterpreted the LUX 
limits accordingly~\cite{Buchmueller:2014yoa,Savage:2015xta,Chala:2015ama}.

We will use the limits on the spin-dependent WIMP-neutron scattering cross section 
presented in Ref.~\cite{Savage:2015xta}.
They are very similar to those presented in Ref.~\cite{Buchmueller:2014yoa}. Since 
Xenon, the target material of LUX, has neutron-odd isotopes, 
limits on spin-dependent WIMP-proton scattering are considerably weaker and are 
thus not taken into account. The results are shown in Fig.~\ref{fig:cosmores} (purple curve).

\section{Results and Discussion} \label{sec:results}

\begin{figure}[h!]
\centering
\setlength{\unitlength}{1\textwidth}
\begin{picture}(0.99,0.933)
  \put(0.0,0.0){\includegraphics[width=0.98\textwidth]{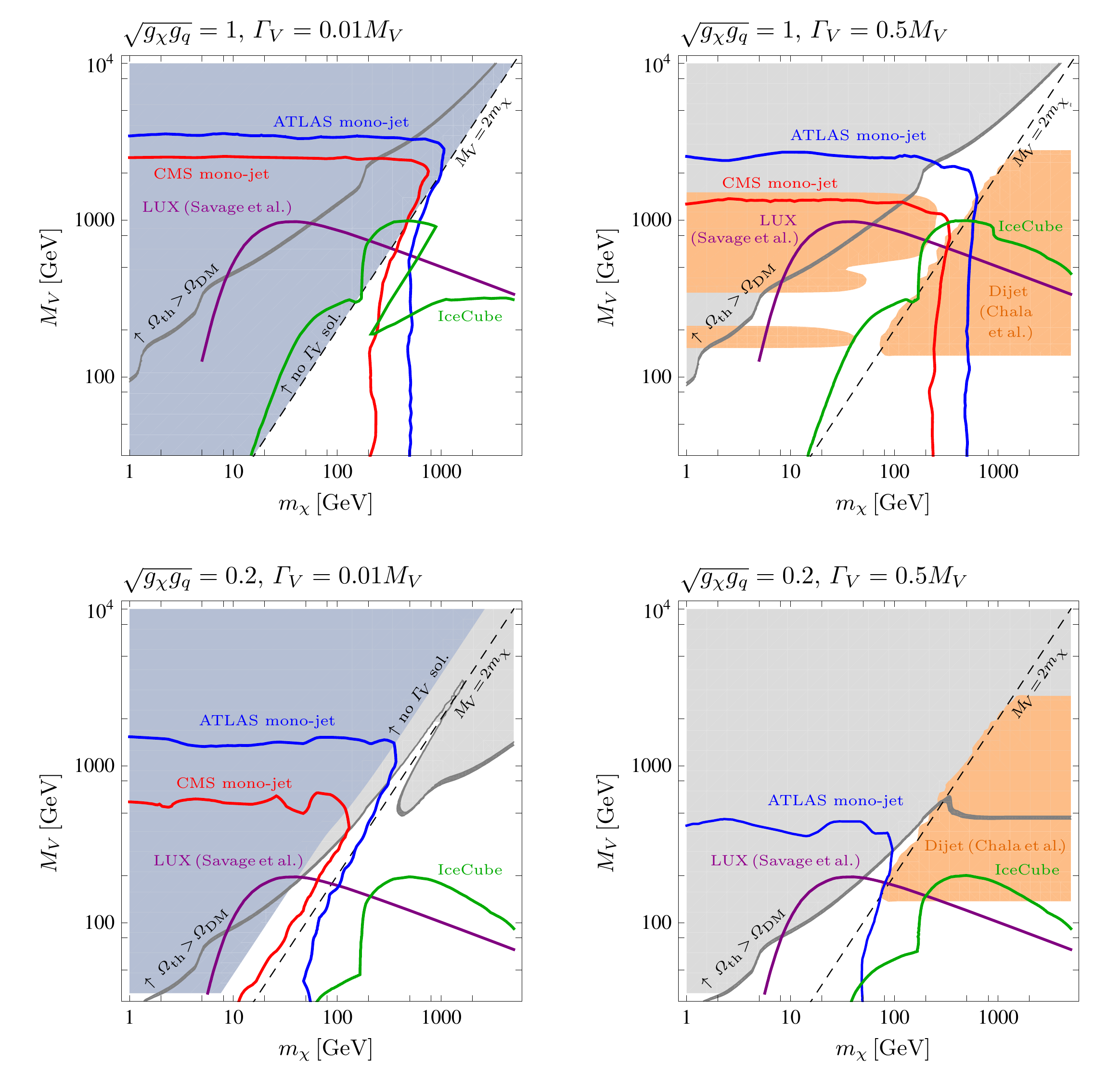}}
\end{picture}
\caption{
Summary of the exclusion limits in the $m_\chi$-$M_{V}$ plane
in four slices of the considered parameter space: 
$\sqrt{g_\chi g_q}=1$, $\Gamma_V=0.01 M_V$ (upper left panel),
$\sqrt{g_\chi g_q}=1$, $\Gamma_V=0.5 M_V$ (upper right panel),
$\sqrt{g_\chi g_q}=0.2$, $\Gamma_V=0.01 M_V$  (lower left panel) and
$\sqrt{g_\chi g_q}=0.2$, $\Gamma_V=0.5 M_V$  (lower right panel).
We show the 95\%\,CL lower exclusion limits from mono-jet searches at ATLAS (blue lines) 
and CMS (red lines) (both in the simplified model interpretation) as well as
limits from searches for resonances in di-jet signatures taken from Ref.~\cite{Chala:2015ama}
(orange shaded regions are excluded).
Furthermore, we show the 90\%\,CL lower exclusion limits from the
IceCube Neutrino Observatory (green line) and from LUX derived from the limits presented
in Ref.~\cite{Savage:2015xta} (purple line).
The dark grey shaded band denotes the region where the relic density matches the 
dark matter density within $\pm10\%$. In the light-grey shaded
region above it, the dark matter is over-produced. 
The blue shaded region in the left panels do not allow for
a consistent solution for the mediator width as a function of 
$M_V,m_\chi,\sqrt{g_\chi g_q}$ within the model.}
\label{fig:mainres}
\end{figure}

In this section we finally summarize the constraints on our model from indirect, direct and 
collider searches. Fig.~\ref{fig:mainres} presents limits on the dark matter and mediator 
masses for four different choices of the coupling strength, $\sqrt{g_\chi g_q}$, and 
the mediator width, $\Gamma_V$.  As discussed in Sec.~\ref{sec:model}, not all 
combinations of the parameters $m_\chi, M_V, \Gamma_V$ and $\sqrt{g_\chi g_q}$ are 
viable within our model. For small width, a large range of the parameter space, marked 
in shaded blue in the left panels of Fig.~\ref{fig:mainres}, is not allowed. For a large width,  
$\Gamma_V = 0.5\, M_V$,  on the other hand, the whole parameter region is allowed.

The model provides the correct dark matter relic density only for the parameter region within 
the thin grey band. The shaded grey area in Fig.~\ref{fig:mainres} indicates the parameter 
region leading to an overproduction of dark matter within the model.
Note, however, that the relic density constraint can be softened
significantly if we assume an extended particle spectrum that leads to co-annihilation effects 
(which can give access to the region above the thermal relic-band) or to additional -- non-thermal -- 
contributions to the dark matter production (which can give access to the region below the thermal 
relic-band).

The limits from the ATLAS and CMS mono-jet searches, obtained within the simplified model, 
are particularly relevant for small $m_\chi$. The ATLAS mono-jet analysis (blue solid curve) 
is the most constraining search for $M_V>2m_\chi$ in the whole  parameter space considered. 
In particular, this search constrains most strongly the parameter space where the
relic density from thermal freeze-out agrees with the measured dark matter density,
\textit{cf.}\ the dark grey band that represents the parameter region with 
$\Omega_\text{th} h^2 = 0.1199\pm 0.012$. The LHC limits depend upon the width of the mediator. 
This dependence is particularly pronounced for $\sqrt{g_\chi g_q}=0.2$, where LHC searches 
probe mediator masses $M_V\lesssim1\,$TeV. Note that this mass range is of the order of the 
typical scattering energies of mono-jet searches with large MET, so that an EFT description would 
not be reliable. The sensitivity of the LHC searches is higher for smaller width as the cross section 
is enhanced through on-shell mediator production.

As discussed in Sec.~\ref{sec:dijet} we also show the limits from di-jet searches as obtained in 
Ref.~\cite{Chala:2015ama}. These constraints are particularly important for large mediator width 
(right panels), because a larger width requires a larger mediator-quark coupling $g_q$, and thus 
a larger cross section.

The direct and indirect dark matter searches by the LUX and IceCube experiments, respectively, 
are particularly relevant for intermediate and large dark matter masses. The IceCube limits, 
specifically, show a maximal sensitivity for dark matter masses around $200\!\!-\!\!1000$\,GeV. 
This results from two effects:  As the sensitivity for annihilation into $b\bar b$ is significantly 
lower than for $t \bar t$ (\textit{cf.} Fig.~\ref{fig:xsSDlimICcomp}) there is a significant strengthening 
of the IceCube limits around the top threshold. For very large $m_\chi$, on the other hand, 
the capture efficiencies and thus the cross section limits decrease and the IceCube search loses 
sensitivity. Furthermore, for $M_V<m_\chi/2$, the annihilation into a pair of mediators is allowed 
(via a $t$-channel $\chi$), which again provides a lower model rejection than $t \bar t$. Hence, the 
limit depends strongly on the relative importance of annihilation into $t \bar t$ and $VV$. 
For $m_\chi>M_V$ and $\sqrt{g_\chi g_q}=1$, the limit is considerably weaker for a small 
mediator width, as a smaller width requires a smaller $g_q$ and hence a larger $g_\chi$ (note that 
the cross section for annihilation into mediators scales like $g_\chi^4$).

\section{Conclusion} \label{sec:conclude}

The complementarity of direct, indirect and collider searches for dark matter can be exploited 
in the framework of simplified models, where dark matter and its experimental signatures are
described with a minimal amount of new particles, interactions and model parameters. We have 
considered a simplified model with a Majorana fermion dark matter particle and an axial-vector 
mediator with universal couplings to SM quarks. Such a model leads to spin-dependent 
interactions and can thus be probed by IceCube in the search for neutrinos from dark matter 
annihilation in the Sun. 

We have focused on a re-interpretation of IceCube limits on the dark matter annihilation rate 
within our model, and have obtained substantial constraints on the dark matter and mediator 
masses. Furthermore, we have derived new constraints on our model from recent LHC 
mono-jet searches, and have analyzed in some detail the differences between interpretations 
of LHC searches within simplified models and effective field theories. 

We find that the limits from the ATLAS and CMS mono-jet searches are particularly relevant 
for small dark matter masses. They exclude mediator masses $M_V \lesssim 1$\,TeV, depending 
in detail on the size of the couplings and the mediator width. The indirect searches for dark matter 
annihilation in the Sun by IceCube  probe intermediate and large dark matter masses and show 
a maximal sensitivity for masses $m_\chi \simeq 200\!\!-\!\!1000$\,GeV. In this region, 
the dark matter capture efficiencies in the Sun are still sizeable, and dark matter annihilation 
is predominantly into top quarks, leading to more highly energetic neutrinos and thus a higher 
neutrino detection efficiency with IceCube. 

We have also computed the relic density within our model, and have combined the LHC 
mono-jet and IceCube limits with constraints from direct detection and the collider search 
for di-jet resonances. We have found a striking complementarity of the different experimental 
approaches, which probe particular and often distinct regions of the model parameter 
space. Thus, the combination of future collider, indirect and direct searches for dark matter 
will allow a comprehensive test of minimal dark matter models.  

\section*{Acknowledgments}

We would like to thank Chiara Arina, Pavel Gretskov, Carlos de los Heros, Kerstin Hoepfner, 
Alexander Knochel, Manfred Krauss, Lennart Oymanns, Mohamed Rameez, Carsten Rott,
Pat Scott and Jory Sonnenveld for helpful discussions. We acknowledge support by the German 
Research Foundation DFG through the research unit ``New physics at the LHC'', the 
Helmholtz Alliance for Astroparticle Physics and the German Federal Ministry of Education 
and Research BMBF.

\addcontentsline{toc}{section}{References}
\bibliographystyle{utphys}
\bibliography{LHC_DM}

\end{document}